\documentclass[a4paper, 10pt, twoside]{article}
\title{The bitensor algebra through operads}
\author{Pepijn van der Laan and Ieke Moerdijk}
\date{2nd April 2003}
\usepackage[english]{babel}
\usepackage{amsmath, amssymb,theorem}
\usepackage[all]{xy}
\usepackage{epsfig,color}
\xyoption{2cell}
\UseAllTwocells
\UseComputerModernTips
\usepackage{fullpage}
\theoremheaderfont{\scshape}
\theorembodyfont{\itshape}
\theoremstyle{change}
\newtheorem{Tm}{Theorem}[section]

\newtheorem{Cr}[Tm]{Corollary}

\theorembodyfont{\rmfamily}

\newtheorem{Df}[Tm]{Definition}

\newcommand{\be}{\begin{equation}}
\newcommand{\ee}{\end{equation}}
\renewcommand{\phi}{\varphi}

\newcommand{\eps}{\varepsilon}

\newcommand{\id}{\text{id}}

\begin{document}
\maketitle
\abstract{
This note shows that the construction of the bitensor algebra of a
bialgebra (Brouder \cite{B}) can be understood 
in terms of the construction of the cooperad of a bialgebra
(Berger-Moerdijk \cite{BM}), and the construction of a bialgebra from a cooperad
(Van der Laan \cite{vdL}, Frabetti-Van der Laan \cite{FvdL}).}
\section{Introduction}
Throughout this note we will work in the category of vector spaces
over a field $k$. At some point we need to restrict to 1-reduced
(non-$\Sigma$) cooperads. A 1-reduced (non-$\Sigma$) cooperad is a
(non-$\Sigma$) cooperad $C$ such that $C(1) = k$, and $C(0) =0$. For a
collection $C$ we will extensively use the grading on the total space
$\bigoplus_nC(n)$ defined by $\left(\bigoplus_nC(n)\right)^m =C(m+1)$.  
Throughout this note we denote by
$T$ the unital free associative algebra functor, and  by $T^+$ the
non-unital free associative algebra functor. Similarly, $S$
(resp. $S^+$) will denote the free unital (reps. non-unital)
commutative  algebra functor. On pointed vector spaces
(i.e. vector spaces $V$ together with a non-zero linear map
$u:k\longrightarrow V$), we define the pointed tensor algebra $T_*V$ as
the quotient of $TV$ by the ideal generated by $(1-u(1))$. Similarly,
we define $S_*V$ as the quotient of $SV$ by the ideal generated by
$(1-u(1))$. 
Let $B$ be a bialgebra with comultiplication $\Delta$. To $B$
we associate the opposite bialgebra 
$B^{\text{op}}$ with comultiplication $\Delta^{\text{op}} = s
\circ\Delta$, where $s$ is the symmetry of the tensor product.  
We will use that $B$ is a Hopf algebra iff $B^{\text{op}}$ is a Hopf
algebra.
\section{Constructions}
\begin{Df}[\cite{FvdL,vdL}]
Let $C$ be a non-$\Sigma$ cooperad with cocomposition given
by maps 
\[
\gamma^*:C(n) \longrightarrow \bigoplus_{k,n_1+\ldots+n_k = n}C(k)
\otimes (C(n_1)\otimes\ldots\otimes C(n_k)).
\]
Define the graded bialgebra $B_C$ to be the tensor algebra
$T(\bigoplus_nC(n))$ on the total space of $C$ as a graded
algebra (w.r.t. the usual grading of the total space). Use the natural
inclusions 
\[
i_1:C(k)\longrightarrow T(\bigoplus_mC(m)), \quad\text{and}\quad
i_2:C(n_1)\otimes\ldots\otimes C(n_k)\longrightarrow T(\bigoplus_mC(m))
\] 
to define the coproduct on generators as
\[
\Delta = (i_1\otimes i_2)\circ\gamma^*
\]
Extend as an algebra morphism. Define a counit $\eps$ such
that $\eps$ vanishes on generators of degree $\neq 0$, and
$\eps|_{C(1)} = \eps_C$, the coidentity of $C$. Again extend this as an
algebra morphism. Coassociativity and counitality are immediate from
the corresponding properties for $C$. 
\end{Df}
\begin{Df}[\cite{FvdL,vdL}]
Let $C$ be a cooperad (with $S_n$-action on $C(n)$). The
bialgebra structure defined above descends 
to the symmetric algebra $S(\bigoplus_nC(n)_{S_n})$ on the
total space of coinvariants of $C$. This bialgebra is denoted $\bar
B_C$. To see that this bialgebra is welldefined, just note that for a
cooperad we can write the cocomposition as 
\[
\gamma^*:C(n) \longrightarrow \bigoplus_k(\bigoplus_{n_1+\ldots+n_k = n}C(k)
\otimes (C(n_1)\otimes\ldots\otimes C(n_k)))^{S_k}.
\]
\end{Df}

\begin{Df}[\cite{FvdL,vdL}]
Let $C$ be a 1-reduced non-$\Sigma$ cooperad. Define the Hopf
algebra $H_C$ to be $T_*(\bigoplus_nC(n))$ as an algebra with respect
to the basepoint given by the inclusion of $C(1)=k$. The coalgebra
structure is induced by the coalgebra structure on $B_C$. Since $H_C$
is a connected graded bialgebra, the existence of an antipode is
assured.
\end{Df}
\begin{Df}[\cite{FvdL,vdL}]
If $C$ be a 1-reduced cooperad, the Hopf algebra structure descends to
the pointed symmetric algebra $S_*(\bigoplus_nC(n)^{S_n})$ on the total
space of invariants of $C$. This Hopf algebra is denoted $\bar H_C$.
\end{Df} 
\begin{Df}[\cite{BM}]
Let $B$ be a bialgebra. Let $C_B$ the non$\Sigma$ cooperad $C_B(n)=
B^{\otimes n}$ (for $n\geq 1$), with the cocomposition $\gamma^*$
defined on summands by the diagram below.
\[
\xymatrix{
B^{\otimes n} \ar@{.>}[r]^{\gamma^*\qquad}\ar[d]_{\Delta} & B^{\otimes k}
\otimes (B^{\otimes n_1}\otimes\ldots \otimes B^{\otimes n_k})\\
B^{\otimes n}\otimes B^{\otimes n} \ar@{=}[r] &(B^{\otimes
n_1}\otimes\ldots \otimes B^{\otimes n_k}) \otimes (B^{\otimes
n_1}\otimes\ldots \otimes B^{\otimes n_k}),
\ar[u]_{(\mu_1\otimes\ldots\otimes \mu_k)\otimes \id}
}
\]
where $\Delta$ is the coproduct of $B^{\otimes n}$, and
$\mu_i:B^{\otimes n_i}\longrightarrow B$ is the multiplication of the
algebra $B$.  
In Sweedler's notation one can write the cocomposition $\gamma^*$ of $C_B$
on a generator $(x^1,\ldots,x^n)\in C_B(n)$ as
\[
\begin{split}
\gamma^*&(x^1,\ldots,x^n) =\\ &
\sum\sum(x^1_{(1)}\star\ldots\star
  x^{n_1}_{(1)},\ldots,x^{n-n_k+1}_{(1)}\star\ldots\star x^{n}_{(1)}) \otimes
  ((x^1_{(2)},\ldots,x^{n_1}_{(2)})\otimes
  \ldots\otimes(x^{n-n_k+1}_{(2)},\ldots,x^n_{(2)})),
\end{split}
\]
where the first sum is over all $k$ and all partitions $n = n_1+
\ldots +n _k$, and the second sum is the sum of the Sweedler notation, and  
where $\star$ denotes the product of $B$. Note that the unit of $B$ makes
$C_B$ a coaugmented non-$\Sigma$ cooperad. If the bialgebra $B$ is
commutative, then $C_B$ is a cooperad.
\end{Df}
\section{The Bitensor Algebra}
Recall the bialgebras $T(T^+(B))$ and $S(S^+(B))$ (Brouder \cite{B}).
Comparing the equation for $\gamma^*$ in Sweedler's notation with
Brouder's formulas yields the following result.
\begin{Cr}\label{Pp:Brouder}
Let $B$ be a bialgebra. The bialgebra $T(T^+(B))$
is isomorphic to the opposite bialgebra of $B_{C_B}$.
If $B$ is commutative, the bialgebra $S(S^+(B))$ is isomorphic to the
opposite bialgebra of $\bar B_{C_B}$.
\end{Cr}
Let $C$ be a coaugmented cooperad. Then the collection $C^{>1}$
defined by $C^{>1}(1)= k$ and $C^{>1}(n) = C(n)$ for $n>1$ is a
1-reduced cooperad with cocomposition induced by cocomposition in
$C$.
\begin{Cr}
The Pinter Hopf algebra associated to $T(T^+(B))$ (cf. \cite{B}
for terminology) is isomorphic to the opposite Hopf algebra of
$H_{C^{>1}_B}$, and for $B$ commutative the Pinter Hopf algebra
associated to $S(S^+(B))$ is isomorphic to the opposite Hopf algebra
of $\bar H_{C^{>1}_B}$. 
\end{Cr}

\begin{center}
\textsc{Pepijn van der Laan} (vdlaan@math.uu.nl) and \textsc{Ieke Moerdijk}
(moerdijk@math.uu.nl)\\
Mathematisch Instituut, Universiteit Utrecht, P.O.Box 80.010, 3508TA
Utrecht, The Netherlands
\end{center}

\begin{thebibliography}{FvdL}
\bibitem{BM} C. Berger and I. Moerdijk - Axiomatic homotopy theory for
  operads. Preprint \texttt{math.AT/0206094}, 2002.
\bibitem{B} C. Brouder and W. Schmitt - Quantum groups and quantum
  field theory III. Renormalisation. Preprint
  \texttt{arXiv:hep-th/0210097}, 2002.
\bibitem{FvdL} A. Frabetti and P.P.I. van der Laan - Groups and Hopf
  algebras from Operads. Work in progress.
\bibitem{vdL} P.P.I. van der Laan - Ph. D. Thesis, 2003.
\end{thebibliography}
\end{document}